\documentclass[11pt]{article}

\usepackage{amsfonts,amssymb,chet,dsfont,graphicx,mathrsfs,pgfplots,tikz}

\usetikzlibrary{decorations.markings,decorations.pathmorphing,arrows}

\tikzset{cross/.style={path picture={
      \draw[black]
            (path picture bounding box.south east) --
            (path picture bounding box.north west)
            (path picture bounding box.south west) --
            (path picture bounding box.north east);}}}

\title{Holomorphic Scalar Portals and the OPE}

\author{Jean-Fran\c{c}ois Fortin\email{jean-francois.fortin@phy.ulaval.ca} and Shanny Pelchat-Voyer\email{shanny.pelchat-voyer.1@ulaval.ca}}

\affiliation{
D\'epartement de Physique, de G\'enie Physique et d'Optique,\\Universit\'e Laval, Qu\'ebec, QC G1V 0A6, Canada
}

\abstract{Visible-sector SUSY-breaking effects are computed in terms of hidden-sector correlation functions for generic holomorphic scalar portals.  The solutions, which are valid irrespective of the hidden-sector dynamics, are approximated with the help of the operator product expansion (OPE).  Indeed, for theories with superconformal symmetry at high energy, the superconformal OPE formalism can be used to disentangle the high-energy dynamics, encoded in the OPE coefficients, from the low-energy dynamics of the SUSY-breaking vacuum expectation values.  A systematic method is proposed to compute the OPE coefficients, using relations between correlation functions of superfields and correlation functions of their quasi-primary component fields.  The method, which is quite general, could be useful in building models of gauge- or gravity-mediated SUSY breaking and in analysing the viability of such models in a systematic way.}

\date{September 2017} 

\begin{document}

\maketitle



\section{Introduction}\label{SIntro}

Supersymmetry (SUSY) is one of the best theoretical ideas for physics beyond the Standard Model (SM).  Although the SM explains very well all particle physics down to the smallest scales presently probed at the Large Hadron Collider (LHC), the SM cannot explain several physical observations.  For example, in the SM neutrinos are massless and thus the SM cannot account for the observed neutrino oscillations \cite{Olive:2016xmw}.  Moreover, the SM does not include gravity, the weakest and probably most mysterious force.  Of the several ideas put forward to enlarge the SM and explain its shortcomings, SUSY is the most theoretically-motivated one.  Indeed, SUSY is particular since it is based on an extra symmetry principle, which dictates that bosons and fermions transform together in appropriate SUSic representations.  Moreover, when SUSY is gauged, gravity is automatically included.

If SUSY were not broken, the superpartners, \textit{i.e.} the remaining fields in the SUSic representations of the SM fields, would have the same masses than their SM counterparts.  This property implies that quantum corrections to fundamental scalar fields cancel exactly in SUSic theories, stabilizing their masses.  Since the superpartners do not have the same masses than the SM fields, SUSY must be broken.  The scale of SUSY breaking dictates the size of the quantum corrections to fundamental scalar fields.  Hence, for this remarkable property low-scale SUSY breaking has been one important contender to explain the stability of the Higgs boson mass and thus the smallness of the electroweak scale.  However, superpartners have still not been observed at the LHC, therefore the scale of SUSY breaking has been pushed higher than expected by naturalness arguments and the SUSY solution to the hierarchy problem is not as convincing as it once was.

To preserve the benefits of SUSY with respect to fundamental scalars, SUSY breaking must be spontaneous.  Moreover, to explain the smallness of the Higgs mass, SUSY breaking should be implemented by a dynamical mechanism generating naturally small numbers.  There are obviously several ways to break SUSY spontaneously and dynamically, therefore there is an abundance of possible SUSY-breaking effects on the Minimal SUSic SM (MSSM).  To understand the general features of SUSY and the general SUSY-breaking effects on the MSSM, a unified formalism describing the effects of SUSY breaking on the visible sector irrespective of the particular SUSY-breaking hidden sector is of great interest.

Such a formalism exists for gauge mediation, dubbed general gauge mediation (GGM) \cite{Meade:2008wd,Buican:2008ws}.  For gauge-mediated SUSY-breaking models, SUSY breaking is communicated from the hidden sector to the MSSM through extra fields charged under the MSSM gauge group.  GGM allows the computation of the corrections to the visible-sector parameters in terms of correlation functions of hidden-sector fields.  Therefore, the hidden-sector dynamics, encoded in the correlation functions, does not have to be weakly coupled.  Visible-sector corrections from strongly-coupled hidden sectors are given by the same equations.  GGM thus leads to a unified formalism for gauge-mediated SUSY-breaking models where the knowledge of the correlation functions of hidden-sector fields gives all the corrections to the visible-sector parameters.

A careful analysis of the correlation functions is sufficient to determine their parametric dependence on the SUSY-breaking vacuum expectation values (vevs).  Hence, it is possible to compute the visible-sector corrections parametric dependence on the SUSY-breaking vevs.  However, to obtain more quantitative results for the visible-sector corrections, a more powerful method is necessary.  To gain insights into the corrections to the visible-sector parameters for weakly-coupled as well as strongly-coupled hidden sectors, it is possible to express the two-point correlation functions of hidden fields in terms of the operator product expansion (OPE).  The OPE expresses the non-local product of two fields in terms of an infinite sum of local fields on which OPE coefficients times non-local differential operators act.  The OPE coefficients encode the short-distance physics while the differential operators are completely fixed by conformal invariance.  In a conformal field theory (CFT), the OPE is a convergent expansion valid inside correlation functions as long as no other fields are closer to the two fields in the OPE.  In a quantum field theory (QFT), the OPE is an asymptotic expansion valid in the short-distance limit.  Hence, it is possible to approximate the correlation functions of hidden fields appearing in the corrections to the visible-sector parameters with the OPE irrespective of the hidden-sector dynamics.  Moreover, the OPE disentangles the short-distance physics encoded in the OPE coefficients, from the long-distance physics described by the vevs of the fields.  Hence, when the UV physics is under control, as for asymptotically-safe QFTs, the OPE coefficients are computable and all the strongly-coupled effects of the hidden sectors are taken into account by the vevs of the fields.  When SUSY is considered, UV physics is actually described by a superconformal field theory (SCFT).  Indeed, for spontaneous SUSY breaking, the short-distance physics is independent of the SUSY breaking and the OPE techniques can be implemented for SCFTs instead.  The larger superconformal invariance implies relations between quasi-primary components fields of the same superfield.  Hence the OPE coefficients of different quasi-primary components fields, and thus the corrections to the visible-sector parameters, are related to each other.

From the general SCFT formalism of \cite{Osborn:1998qu}, the OPE techniques \cite{Fortin:2011nq} have been used successfully for GGM (see, for example, \cite{Fortin:2011ad,Fortin:2012tp}).  However, amongst other things, GGM in its simplest form does not directly provide a solution to the $\mu$ problem, although a straightforward extension of GGM with extra direct couplings between the hidden sector and the Higgs sector can address it \cite{Komargodski:2008ax}.  Starting again from the general SCFT formalism of \cite{Osborn:1998qu}, the OPE techniques \cite{Poland:2010wg} have been applied to a particular implementation of this extended definition of GGM in \cite{Kumar:2014uxa}.  In all cases, the OPE techniques provide a unified formalism for the visible-sector corrections and bring a new perspective on some generic features observed in specific models.  Moreover, the knowledge brought by the OPE techniques might be useful for new model-building avenues.

In this paper, the unified formalism of GGM is extended to holomorphic scalar portals, \textit{i.e.} to all direct couplings in the superpotential between hidden-sector scalar chiral superfields and visible-sector scalar chiral superfields.  The corresponding corrections to the visible-sector parameters are expressed in terms of one- and two-point correlation functions of hidden-sector fields and the OPE techniques are used to obtain approximate results valid for any type of hidden-sector dynamics.  This paper thus generalizes the work of \cite{Komargodski:2008ax} by including more general superpotential couplings between the hidden sector and the visible sector, as well as the work of \cite{Kumar:2014uxa} by developing the OPE techniques for all models described in \cite{Komargodski:2008ax} and more.

As such, this paper can be seen as a step towards general gravity mediation.  Indeed, in gravity-mediated SUSY-breaking models SUSY breaking is transmitted to the visible sector by Planck-suppressed non-renormalizable (K\"ahler-potential and superpotential) couplings between hidden-sector and visible-sector superfields.  Since such couplings are always present, gravity mediation is always relevant although its importance with respect to other types of mediation, like gauge mediation, depends on the different scales under consideration.

This paper\footnote{The notation from Wess \& Bagger \cite{Wess:1992cp} is used throughout, taking into account factors of $i$ as explained in \cite{Fortin:2011nq}.} is organized as follows: Section \ref{SHV} sets the framework and computes the corrections to the visible-sector parameters in terms of one- and two-point correlation functions of hidden-sector fields.  In section \ref{SOPE} the OPE is introduced and its computation in terms of two- and three-point correlation functions is presented.  To proceed, the three-point correlation functions of quasi-primary component fields are obtained from the known three-point correlation functions of superfields.  Section \ref{SOPEResults} presents the relevant OPEs for all the two-point correlation functions of hidden fields appearing in the corrections to the visible-sector parameters.  In section \ref{SResults} the visible-sector corrections are computed from the results of the previous sections and dispersion relations.  General comments on the corrections to the visible-sector parameters are also made.  Finally, a discussion and a conclusion are presented in section \ref{SConclusion}.


\section{From Hidden Sector to Visible Sector}\label{SHV}

This section fixes the notation and computes the corrections to the visible-sector effective theory in terms of hidden-sector correlation functions generated by integrating out the hidden-sector degrees of freedom.

\subsection{High-energy Degrees of Freedom}

The degrees of freedom at high energy are divided in two sectors: the visible sector and the SUSY-breaking hidden sector.  The visible sector is assumed to be weakly coupled with a canonical K\"ahler potential and a renormalizable superpotential.  The visible-sector scalar chiral superfields, denoted by $\Phi_a$, thus have the following superpotential,
\eqn{\mathcal{W}_\text{V}=\frac{1}{6}y^{abc}\Phi_a\Phi_b\Phi_c+\frac{1}{2}M^{ab}\Phi_a\Phi_b+L^a\Phi_a,}[EqnWV]
where the coupling constants are completely symmetric.

The hidden sector is left unconstrained.  However, the couplings between the hidden sector and the visible sector are assumed to be of the form
\eqn{\mathcal{W}_\text{V$\otimes$H}=\frac{1}{2}h^{abi}\Phi_a\Phi_b\mathcal{O}_i+g^{ai}\Phi_a\mathcal{O}_i,}[EqnWVH]
where the $\mathcal{O}_i$ are hidden-sector scalar chiral superfields.  The coupling constants $h^{abi}=h^{bai}$ and $g^{ai}$ are assumed perturbative.\footnote{More precisely, the dimensionless coupling constants $h^{abi}(\mu)\equiv h^{abi}\mu^{\Delta_i-1}$ and $g^{ai}(\mu)\equiv g^{ai}\mu^{\Delta_i-2}$, where $\mu$ is the renormalization scale, are smaller than one.}

Integrating out the high-energy degrees of freedom of the hidden sector generates corrections to the visible-sector parameters \eqref{EqnWV} due to \eqref{EqnWVH},
\eqna{
\delta\mathcal{W}_\text{V}&=\frac{1}{6}\delta y^{abc}\Phi_a\Phi_b\Phi_c+\frac{1}{2}\delta M^{ab}\Phi_a\Phi_b+\delta L^a\Phi_a,\\
\mathcal{L}_\text{V,soft}&=-\left[\frac{1}{6}a^{abc}\phi_a\phi_b\phi_c+\frac{1}{2}b^{ab}\phi_a\phi_b+t^a\phi_a+u^a_{\phantom{a}b}\phi_aF^{b\dagger}+v_aF^{a\dagger}+\text{h.c.}\right]-(m^2)^a_{\phantom{a}b}\phi_a\phi^{b\dagger},\\
\mathcal{L}_\text{V,other}&=\delta Z^a_{\phantom{a}b}F_aF^{b\dagger}-\frac{1}{4}\alpha^{ab}_{\phantom{ab}cd}\phi_a\phi_b\phi^{c\dagger}\phi^{d\dagger}-\left[\frac{1}{24}\alpha^{abcd}\phi_a\phi_b\phi_c\phi_d+\frac{1}{2}\beta^a_{\phantom{a}bc}\phi_a\phi^{b\dagger}\phi^{c\dagger}\right.\\
&\left.\phantom{=}+\frac{1}{2}\gamma^{abc}\phi_a\psi_b\psi_c+\frac{1}{2}\gamma^a_{\phantom{a}bc}\phi_a\bar{\psi}^b\bar{\psi}^c+\frac{1}{2}\lambda^{ab}F_aF_b+\frac{1}{2}\rho^{abc}\phi_a\phi_bF_c+\frac{1}{2}\rho^{ab}_{\phantom{ab}c}\phi_a\phi_bF^{c\dagger}\right.\\
&\left.\phantom{=}+\rho^a_{\phantom{a}bc}\phi_a\phi^{b\dagger}F^{c\dagger}+\zeta^{ab}\phi_aF_b+\text{h.c}\right]-\Lambda^4.
}[EqndV]
Here, only the renormalizable corrections to the visible-sector parameters are shown and the coupling-constant corrections have obvious symmetry properties.

\subsection{Low-energy Degrees of Freedom}

The degrees of freedom at low energy are the visible-sector superfields, however the visible-sector Lagrangian is corrected as mentioned above.  Since the coupling constants mixing the hidden sector with the visible sector are perturbative, the corrections can be expressed as expansions in the coupling constants $h^{abi}$ and $g^{ai}$.  At quadratic order in the coupling constants, one loop and zero momentum, the corrections to the superpotential \eqref{EqndV} using the expansion in components $\mathcal{O}=e^{i\theta Q+i\bar{\theta}\bar{Q}}O$ are given by
\eqna{
\delta y^{abc}&=-\frac{3i}{4}h^{(ab|i}g^{c)j}\int d^4x\langle Q^\alpha O_i(x)Q_\alpha O_j(0)\rangle,\\
\delta M^{ab}&=h^{abi}\langle O_i(0)\rangle-\frac{i}{4}g^{(a|i}g^{b)j}\int d^4x\langle Q^\alpha O_i(x)Q_\alpha O_j(0)\rangle,\\
\delta L^a&=g^{ai}\langle O_i(0)\rangle,
}[EqndWV]
where the parenthesis denote properly-normalized symmetrization, \textit{e.g.} $g^{(a|i}g^{b)j}=(g^{ai}g^{bj}+g^{bi}g^{aj})/2$.  As expected the corrections to the visible-sector superpotential \eqref{EqndWV} do not vanish if SUSY is unbroken.  The soft Lagrangian parameters \eqref{EqndV} are generated as
\eqna{
a^{abc}&=-\frac{3i}{16}h^{(ab|i}g^{c)j}\int d^4x\langle Q^2O_i(x)Q^2O_j(0)\rangle,\\
b^{ab}&=-\frac{1}{4}h^{abi}\langle Q^2O_i(0)\rangle-\frac{i}{16}g^{(a|i}g^{b)j}\int d^4x\langle Q^2O_i(x)Q^2O_j(0)\rangle,\\
t^a&=-\frac{1}{4}g^{ai}\langle Q^2O_i(0)\rangle+\frac{1}{16}h^{aci}g_b^{\phantom{b}j}\int d^4x\,\delta^{(4)}(x)\frac{1}{\partial^2}\langle Q^2\bar{Q}^2[O_i(x)O_j^\dagger(0)]\rangle,\\
u^a_{\phantom{a}b}&=-\frac{i}{4}g^{ai}g_b^{\phantom{b}j}\int d^4x\langle Q^2O_i(x)O_j^\dagger(0)\rangle+\frac{1}{4}h^{aci}h_{bc}^{\phantom{bc}j}\int d^4x\,\delta^{(4)}(x)\frac{1}{\partial^2}\langle Q^2O_i(x)O_j^\dagger(0)\rangle,\\
v_a&=\frac{1}{4}g^{bi}h_{ab}^{\phantom{ab}j}\int d^4x\,\delta^{(4)}(x)\frac{1}{\partial^2}\langle Q^2O_i(x)O_j^\dagger(0)\rangle,\\
(m^2)^a_{\phantom{a}b}&=-\frac{i}{16}g^{ai}g_b^{\phantom{b}j}\int d^4x\langle Q^2O_i(x)\bar{Q}^2O_j^\dagger(0)\rangle+\frac{1}{16}h^{aci}h_{bc}^{\phantom{bc}j}\int d^4x\,\delta^{(4)}(x)\frac{1}{\partial^2}\langle Q^2\bar{Q}^2[O_i(x)O_j^\dagger(0)]\rangle,
}[EqndLVs]
while the other contributions \eqref{EqndV} are
\eqna{
\delta Z^a_{\phantom{a}b}&=ig^{ai}g_b^{\phantom{b}j}\int d^4x\langle O_i(x)O_j^\dagger(0)\rangle-h^{aci}h_{bc}^{\phantom{bc}j}\int d^4x\,\delta^{(4)}(x)\frac{1}{\partial^2}\langle O_i(x)O_j^\dagger(0)\rangle,\\
\alpha^{ab}_{\phantom{ab}cd}&=-\frac{i}{16}h^{abi}h_{cd}^{\phantom{cd}j}\int d^4x\langle Q^2O_i(x)\bar{Q}^2O_j^\dagger(0)\rangle,\\
\alpha^{abcd}&=-\frac{3i}{16}h^{(ab|i}h^{cd)j}\int d^4x\langle Q^2O_i(x)Q^2O_j(0)\rangle,\\
\beta^a_{\phantom{a}bc}&=-\frac{i}{16}g^{ai}h_{bc}^{\phantom{bc}j}\int d^4x\langle Q^2O_i(x)\bar{Q}^2O_j^\dagger(0)\rangle,\\
\gamma^{abc}&=\frac{i}{4}g^{ai}h^{bcj}\int d^4x\langle Q^\alpha[Q_\alpha O_i(x)O_j(0)]\rangle,\\
\gamma^a_{\phantom{a}bc}&=\frac{i}{4}g^{ai}h_{bc}^{\phantom{bc}j}\int d^4x\langle Q^2O_i(x)O_j^\dagger(0)\rangle,\\
\lambda^{ab}&=-ig^{(a|i}g^{b)j}\int d^4x\langle O_i(x)O_j(0)\rangle,\\
\rho^{abc}&=-\frac{i}{4}\left(\frac{1}{2}h^{abi}g^{cj}+g^{(a|i}h^{b)cj}\right)\int d^4x\langle Q^\alpha[Q_\alpha O_i(x)O_j(0)]\rangle,\\
\rho^{ab}_{\phantom{ab}c}&=-\frac{i}{4}h^{abi}g_c^{\phantom{c}j}\int d^4x\langle Q^2O_i(x)O_j^\dagger(0)\rangle,\\
\rho^a_{\phantom{a}bc}&=-\frac{i}{4}g^{ai}h_{bc}^{\phantom{bc}j}\int d^4x\langle Q^2O_i(x)O_j^\dagger(0)\rangle,\\
\zeta^{ab}&=-\frac{i}{4}g^{ai}g^{bj}\int d^4x\langle Q^\alpha[Q_\alpha O_i(x)O_j(0)]\rangle,\\
\Lambda^4&=\frac{1}{16}g^{ai}g_a^{\phantom{a}j}\int d^4x\,\delta^{(4)}(x)\frac{1}{\partial^2}\langle Q^2\bar{Q}^2[O_i(x)O_j^\dagger(0)]\rangle.
}[EqndLVo]
Hence the corrections to the visible-sector parameters can be divided into two groups: the corrections proportional to zero-momentum Fourier transforms of the two-point correlation functions, and the corrections where the spacetime integrals include $\delta^{(4)}(x)\frac{1}{\partial^2}$ times the two-point correlation functions.

All the soft Lagrangian parameters vanish when SUSY is unbroken.  This is clear since the two-point correlation functions for soft Lagrangian parameters can be written as $\langle Q(\cdot)\rangle$, as in $\langle Q^2O_i(x)Q^2O_j(0)\rangle=-\langle Q^2[Q^\alpha O_i(x)Q_\alpha O_j(0)]\rangle$, except for $\langle Q^2O_i(x)\bar{Q}^2O_j^\dagger(0)\rangle$.  For this case, one has ($\{Q_\alpha,\bar{Q}_{\dot{\alpha}}\}=2\sigma_{\alpha\dot{\alpha}}^\mu P_\mu$ from the algebra)
\eqn{\langle Q^2O_i(x)\bar{Q}^2O_j^\dagger(0)\rangle=\langle Q^\alpha[Q_\alpha O_i(x)\bar{Q}^2O_j^\dagger(0)]\rangle-4i\partial^{\dot{\alpha}\alpha}\langle Q_\alpha O_i(x)\bar{Q}_{\dot{\alpha}}O_j^\dagger(0)\rangle,}
which vanish at zero momentum when SUSY is unbroken.  Since corrections proportional to $\langle Q^2O_i(x)\bar{Q}^2O_j^\dagger(0)\rangle$ are always zero-momentum Fourier transforms, the latter vanish in the SUSic limit.

Apart from the expected wave-function renormalizations $\delta Z^a_{\phantom{a}b}$ which contribute to the K\"ahler potential, the other contributions also vanish when SUSY is unbroken.  This statement is clear for all contributions except $\lambda^{ab}$.  In this case, if $\langle O_i(x)O_j(0)\rangle$ do not vanish when SUSY is conserved, then their values do not depend on the separation $|x|$ between the two scalar chiral quasi-primary operators \cite{Amati:1988ft}.  Therefore $|x|$ can be taken to infinity and by the cluster-decomposition theorem, $\langle O_i(x)O_j(0)\rangle=\langle O_i\rangle\langle O_j\rangle$.  Hence the individual scalar chiral quasi-primary operators must have non-vanishing vevs.  This can happen only when they are not charged under the visible-sector gauge group.  One concludes that if the scalar chiral quasi-primary operators are charged under the visible-sector gauge group, then the two-point correlation functions $\langle O_i(x)O_j(0)\rangle$ is non-vanishing only when SUSY is broken, as expected.  However, if the scalar chiral quasi-primary operators are not charged under the visible-sector gauge group, then the two-point correlation functions $\langle O_i(x)O_j(0)\rangle$ could be non-zero even when SUSY is preserved, in apparent contradiction with general SUSY-breaking lore.  The methods developed here will give an explanation as to why these possible non-vanishing SUSY-preserving contributions do not appear in $\lambda^{ab}$.

At this order, all the results are expressed in terms of hidden-sector one- and two-point correlation functions.  The relevant two-point correlation functions are
\eqn{
\begin{gathered}
\langle O_i(x)O_j(0)\rangle,\quad\quad\langle Q^\alpha[Q_\alpha O_i(x)O_j(0)]\rangle,\quad\quad\langle Q^\alpha O_i(x)Q_\alpha O_j(0)\rangle,\quad\quad\langle Q^2O_i(x)Q^2O_j(0)\rangle,\\
\langle O_i(x)O_j^\dagger(0)\rangle,\quad\quad\langle Q^2O_i(x)O_j^\dagger(0)\rangle,\quad\quad\langle Q^2\bar{Q}^2[O_i(x)O_j^\dagger(0)]\rangle,\quad\quad\langle Q^2O_i(x)\bar{Q}^2O_j^\dagger(0)\rangle.
\end{gathered}
}[EqnBraOiOjKet]
These two-point correlation functions are computable perturbatively in weakly-coupled hidden sectors.  Unfortunately, perturbation theory is of no help for strongly-coupled hidden sectors.  Moreover, a careful analysis of the correlation functions in the limit of small vacuum energy density with respect to the typical mass scale of the hidden sector only leads to a parametric dependence of the visible-sector corrections in terms of the SUSY-breaking vevs \cite{Komargodski:2008ax}.  Another technique is necessary to obtain more quantitative results.

It is possible to approximate the corrections to the visible sector generically using OPE techniques.  The two-point correlation functions \eqref{EqnBraOiOjKet} thus lead to the following independent two-point functions
\eqn{
\begin{gathered}
O_i(x)O_j(0),\quad\quad Q_\alpha O_i(x)O_j(0),\quad\quad Q^\alpha O_i(x)Q_\alpha O_j(0),\\
O_i(x)O_j^\dagger(0),\quad\quad Q^2O_i(x)\bar{Q}^2O_j^\dagger(0),
\end{gathered}
}[EqnOiOj]
which will have to be expressed in terms of the OPE.

The power of the OPE lies with strongly-coupled theories since complicated low-energy dynamics is completely encoded in vevs.  Moreover, when compared to the small vacuum-energy-density limit, the OPE allows a more quantitative knowledge of the visible-sector corrections since it computes contributions to the visible-sector corrections systematically, leading to exact numbers times proper powers of SUSY-breaking vevs.

Since the resulting vevs must not break Lorentz invariance, the relevant OPEs must be with scalar fields only, apart from the $Q_\alpha O_i(x)O_j(0)$ OPE which must be with spinor fields only.  In terms of superfields, this implies that the supersymmetric OPEs are solely with superfields in the scalar, spinor and vector irreducible representations.


\section{Operator Product Expansion and Correlation Functions}\label{SOPE}

This section reviews the OPE as well as the two- and three-point correlation functions and demonstrate how the OPE coefficients can be extracted from the knowledge of these correlation functions.

\subsection{From Correlation Functions to Operator Product Expansion}

The OPE expresses the non-local product of two fields $O_i(x_1)$ and $O_j(x_2)$ at different spacetime points in terms of an infinite sum of local fields $O_k(x_2)$ on which (non-local) differential operators $\mathcal{D}_{ij}^{\phantom{ij}k}(x_{12},\partial_2)$ act,
\eqn{O_i(x_1)O_j(x_2)=\sum_k\lambda_{ij}^{\phantom{ij}k}\mathcal{D}_{ij}^{\phantom{ij}k}(x_{12},\partial_2)O_k(x_2).}[EqnOPE]
In a CFT, the OPE is exact and the differential operators are completely fixed by conformal invariance.  In a UV asymptotically-safe QFT, it is possible to approximate non-local products of two fields in terms of the OPE.  In this setup, the OPE coefficients $\lambda_{ij}^{\phantom{ij}k}$ are fixed by UV physics while the IR physics is encoded in the vevs of the fields.  Since the OPE coefficients can be easily computed, this approximation is particularly interesting when the QFT is strongly coupled in the IR.  Indeed, it allows computing physical quantities in the strong-coupling regime in terms of calculable OPE coefficients and (strongly-coupled but in principle measurable) vevs.

It is well known that the information encoded in the OPE can be retrieved from the two- and three-point correlation functions.  Indeed, using a convenient (diagonal) basis for the quasi-primary operators of a unitary CFT, conformal invariance implies that the two-point correlation functions are diagonal, \textit{i.e.}
\eqn{\langle O_i(x_1)O_j^\dagger(x_2)\rangle=C_i\frac{I_{i\bar{\textit{\i}}}(x_{12})}{x_{12}^{2\Delta_i}}\delta_{ij},}
where $\Delta$ is the conformal dimension.  Here the exact dependence on $x_{12}^\mu$ is known and is encoded in the function $I_{i\bar{\textit{\i}}}(x)$ which takes care of the irreducible representation of the quasi-primary operators.  The three-point correlation functions are also known and are given by
\eqn{\langle O_i(x_1)O_j(x_2)O_k^\dagger(x_3)\rangle=C_{ij\bar{k}}\frac{I_{ij\bar{k}}(x_{12},x_{13},x_{23})}{x_{12}^{\Delta_i+\Delta_j-\Delta_k}x_{13}^{\Delta_i+\Delta_k-\Delta_j}x_{23}^{\Delta_j+\Delta_k-\Delta_i}},}
where $I_{ij\bar{k}}(x_{12},x_{13},x_{23})$ takes into account the different irreducible representations of the quasi-primary fields such that both sides transform similarly under Lorentz transformations.

From this knowledge, it is clear that there exists a one-to-one correspondence between the three-point correlation functions and the OPE, since using the OPE \eqref{EqnOPE} in the three-point correlation functions leads to
\eqn{\langle O_i(x_1)O_j(x_2)O_k^\dagger(x_3)\rangle=\lambda_{ij}^{\phantom{ij}k}\mathcal{D}_{ij}^{\phantom{ij}k}(x_{12},\partial_2)\langle O_k(x_2)O_k^\dagger(x_3)\rangle.}[EqnOPEC]
Thus the OPE coefficients can be retrieved from the knowledge of the two- and three-point correlation function coefficients $C_i$ and $C_{ij\bar{k}}$ and the differential operators.\footnote{Obviously, different normalizations lead to different OPE coefficients.  For example, differential operators can be normalized as $\mathcal{D}_{ij}^{\phantom{ij}k}(x_{12},\partial_2)=x_{12}^{-\Delta_i-\Delta_j+\Delta_k}(1+\cdots)I_{ij}^{\phantom{ij}k}$ where the ellipses involve (generically still unknown but in principle fixed) partial derivatives.  However, final physical results are always free of normalization issues.}  Moreover, to avoid using the (generically still unknown but in principle fixed) differential operators, a comparison of the two- and three-point correlation functions in the limit $x_{12}^\mu\to0$ is sufficient.

\subsection{From Correlation Functions of Superfields to Correlation Functions of Fields}

The discussion above can be applied directly to SCFTs.  Since the OPE coefficients encode UV physics and the theories of interest here are assumed to become SCFTs in the UV, it is possible to compute the OPE coefficients directly from superconformal correlation functions of superfields.

As explained above, the relevant OPEs are given by \eqref{EqnOiOj} since \eqref{EqnBraOiOjKet} are the two-point correlation functions of interest, therefore the corresponding three-point correlation functions of superfields giving rise to non-vanishing contributions to \eqref{EqnBraOiOjKet} and hence to visible-sector corrections are
\eqn{
\begin{gathered}
\langle\mathcal{O}_i(z_1)\mathcal{O}_j(z_2)\mathcal{O}_{k(0,0)}^\dagger(z_3)\rangle,\quad\quad\langle\mathcal{O}_i(z_1)\mathcal{O}_j(z_2)(\mathcal{O}_{k(1,0)\beta})^\dagger(z_3)\rangle,\quad\quad\langle\mathcal{O}_i(z_1)\mathcal{O}_j(z_2)(\mathcal{O}_{k(1,1)\beta\dot{\beta}})^\dagger(z_3)\rangle,\\
\langle\mathcal{O}_i(z_1)\mathcal{O}_j^\dagger(z_2)\mathcal{O}_{k(0,0)}^\dagger(z_3)\rangle,\quad\quad\langle\mathcal{O}_i(z_1)\mathcal{O}_j^\dagger(z_2)(\mathcal{O}_{k(1,1)\beta\dot{\beta}})^\dagger(z_3)\rangle,
\end{gathered}
}[EqnmOimOj]
where $\mathcal{O}_{(j,\bar{\textit{\j}})}$ is a superfield in the $(j,\bar{\textit{\j}})$ irreducible representation of the Lorentz group.\footnote{Both quasi-primary fields and quasi-primary superfields are denoted by their irreducible representation of the Lorentz group $(j,\bar{\textit{\j}})$ and their conformal dimension $\Delta$.  However, since the superconformal algebra contains an extra charge, quasi-primary superfields are also denoted by their $R$-charge $R$.  It is usually more convenient to use the variables $q=\Delta/2+3R/4$ and $\bar{q}=\Delta/2-3R/4$ for quasi-primary superfields.}  Superfields in other irreducible representations either have vanishing three-point correlation functions [hence the missing spin $1/2$ superfields in the second row of \eqref{EqnmOimOj}] or Lorentz-violating vevs \cite{Kumar:2014uxa}.
\begin{table}[t]
\centering
\resizebox{16cm}{!}{
\begin{tabular}{|c|cccc|}
\hline
$\langle\mathcal{O}_i\mathcal{O}_j\mathcal{O}_{k(0,0)}^\dagger\rangle$ & $\langle O_iO_jO_{k(0,0)}^\dagger\rangle$ & $\langle O_iO_j(Q^2O_{k(0,0)})^\dagger\rangle$ & $\langle O_iO_j(\bar{Q}^2O_{k(0,0)})^\dagger\rangle$ & $\langle O_iO_j[Q^2\bar{Q}^2O_{k(0,0)}]_p^\dagger\rangle$\\\hline
$\varnothing$ & $1$ & $0$ & $0$ & $2^3(c_{13}-c_1c_9)P_3^2$\\
$\theta_3^\gamma\bar{\theta}_3^{\dot{\gamma}}$ & $0$ & $0$ & $0$ & $2^2c_9P_3^{\dot{\gamma}\gamma}$\\
$\theta_3^2$ & $0$ & $0$ & $2^2$ & $0$\\
$\bar{\theta}_3^2$ & $0$ & $2^2$ & $0$ & $0$\\
$\theta_3^2\bar{\theta}_3^2$ & $0$ & $0$ & $0$ & $2^4$\\
\hline
\end{tabular}
}
\caption{Relevant linear combinations for the $O_iO_j$ OPE with spin $0$ superfields $\mathcal{O}_{k(0,0)}$.  For the corresponding $Q^\alpha O_iQ_\alpha O_j$ OPE, the linear combinations consist of the same $\langle\mathcal{O}_i\mathcal{O}_j\mathcal{O}_{k(0,0)}^\dagger\rangle$ components with extra $\theta_1\theta_2$ and all coefficients must be multiplied by $2$.  For the corresponding $O_iO_j^\dagger$ OPE, the linear combinations are exactly the same with the same coefficients.  Finally, for the corresponding $Q^2O_i\bar{Q}^2O_j^\dagger$ OPE, the linear combinations consist of the same $\langle\mathcal{O}_i\mathcal{O}_j^\dagger\mathcal{O}_{k(0,0)}^\dagger\rangle$ components with extra $\theta_1^2\bar{\theta}_2^2$ and all coefficients must be multiplied by $2^4$.}
\label{TabOOO0}
\end{table}
\begin{table}[t]
\centering
\begin{tabular}{|c|cc|}
\hline
$\langle\mathcal{O}_i\mathcal{O}_j\mathcal{O}_{k(0,0)}^\dagger\rangle$ & $\langle Q_\alpha O_iO_j(Q_\beta O_{k(0,0)})^\dagger\rangle$ & $\langle Q_\alpha O_iO_j[\bar{Q}^2Q_\beta O_{k(0,0)}]_p^\dagger\rangle$\\\hline
$\theta_1^\alpha\bar{\theta}_3^{\dot{\beta}}$ & $1$ & $0$\\
$\theta_1^\alpha\theta_{3\gamma}$ & $0$ & $2\bar{c}_5P_{3\gamma\dot{\beta}}$\\
$\theta_1^\alpha\bar{\theta}_3^{\dot{\beta}}\theta_3^2$ & $0$ & $2^2$\\
\hline
\end{tabular}
\caption{Relevant linear combinations for the $Q_\alpha O_iO_j$ OPE with spin $0$ superfields $\mathcal{O}_{k(0,0)}$.}
\label{TabQOOO0}
\end{table}

To obtain the appropriate OPE coefficients of \eqref{EqnOiOj}, it is necessary to derive the three-point correlation functions of quasi-primary component fields from the known three-point correlation functions of superfields \eqref{EqnmOimOj}.  This can be done straightforwardly using the expansion $\mathcal{O}=e^{i\theta Q+i\bar{\theta}\bar{Q}}O$ as in \cite{Li:2014gpa}.  The results are given in tables \ref{TabOOO0} and \ref{TabQOOO0} for spin $0$ superfields $\mathcal{O}_{k(0,0)}$, tables \ref{TabOOO12} and \ref{TabQOOO12} for spin $1/2$ superfields $\mathcal{O}_{k(1,0)}$, and tables \ref{TabOOO1} and \ref{TabQOOO1} for spin $1$ superfields $\mathcal{O}_{k(1,1)}$.

The different coefficients $c_i$ disentangle the contributions of descendants to the quasi-primary fields.  When descendants contribute, the quasi-primary fields are denoted by a subscript ``$p$'' to avoid confusion.  The coefficients $c_i$ are functions of $(j,\bar{\textit{\j}},q,\bar{q})$ and are explicitly given in \cite{Li:2014gpa}.  Here $\bar{c}_i$ is equal to $c_i$ with $(j,\bar{\textit{\j}},q,\bar{q})$ replaced by $(\bar{\textit{\j}},j,\bar{q},q)$.  For more details the reader is referred to \cite{Li:2014gpa}.

It is important to mention that, if their contributions to the two-point correlation functions \eqref{EqnBraOiOjKet} vanish, specific quasi-primary component fields are excluded from tables \ref{TabOOO0} to \ref{TabQOOO1}.  Hence, quasi-primary component fields that vanish due to the action of an extra $Q^\alpha$ charge\footnote{Although quasi-primary component fields are not simply given by the action of charges on the lowest-component fields, the action of an extra charge can be taken as expected since the corrections have vanishing vevs.  For example, $[\bar{Q}^2Q_\alpha O_{(0,0)}]_p=\bar{Q}^2Q_\alpha O_{(0,0)}+2(1-\bar{c}_5)P_{\alpha\dot{\alpha}}\bar{Q}^{\dot{\alpha}}O_{(0,0)}$ so $Q^\alpha[\bar{Q}^2Q_\alpha O_{(0,0)}]_p=[Q^2\bar{Q}^2O_{(0,0)}]_p+\cdots$ where the ellipses are quasi-primary component fields with partial derivatives acting on them.} or that vanish due to shortening conditions\footnote{As described in the next subsection, relevant spin $1/2$ superfields $\mathcal{O}_{k(1,0)}$ are always short operators.} are not included.\footnote{Other specific linear combinations might vanish due to the structure of the three-point correlation functions.}
\begin{table}[t]
\centering
\begin{tabular}{|c|c|}
\hline
$\langle\mathcal{O}_i\mathcal{O}_j(\mathcal{O}_{k(1,0)\beta})^\dagger\rangle$ & $\langle O_iO_j(QO_{k(1,0)})^\dagger\rangle$\\\hline
$\bar{\theta}_{3\dot{\beta}}$ & $-2i$\\
\hline
\end{tabular}
\caption{Relevant linear combination for the $O_iO_j$ OPE with spin $1/2$ superfields $\mathcal{O}_{k(1,0)}$.  For the corresponding $Q^\alpha O_iQ_\alpha O_j$ OPE, the linear combinations consist of the same $\langle\mathcal{O}_i\mathcal{O}_j\mathcal{O}_{k(1,0)}^\dagger\rangle$ components with extra $\theta_1\theta_2$ and all coefficients must be multiplied by $2$.  As described in the next subsection, the quasi-primary field $[\bar{Q}^2QO_{k(1,0)}]_p$ is not included here since it vanishes for the short operator $\mathcal{O}_{k(1,0)}$.}
\label{TabOOO12}
\end{table}
\begin{table}[t]
\centering
\begin{tabular}{|c|c|}
\hline
$\langle\mathcal{O}_i\mathcal{O}_j(\mathcal{O}_{k(1,0)\beta})^\dagger\rangle$ & $\langle Q_\alpha O_iO_j(O_{k(1,0)\beta})^\dagger\rangle$\\\hline
$\theta_1^\alpha$ & $-i$\\
\hline
\end{tabular}
\caption{Relevant linear combination for the $Q_\alpha O_iO_j$ OPE with spin $1/2$ superfields $\mathcal{O}_{k(1,0)}$.  The quasi-primary fields $Q^2O_{k(1,0)\beta}$ and $[Q^2\bar{Q}^2O_{k(1,0)\beta}]_p$ are not included here since their contributions vanish once the remaining $Q^\alpha$ needed in the two-point correlation functions is taken into account.  Moreover, as described in the next subsection, the quasi-primary fields $\bar{Q}^2O_{k(1,0)}$ and $[Q^2\bar{Q}^2O_{k(1,0)\beta}]_p$ are not included here since they vanish for the short operator $\mathcal{O}_{k(1,0)}$.}
\label{TabQOOO12}
\end{table}

For example, from tables \ref{TabOOO0} and \ref{TabOOO1}, it is straightforward to express the following three-point correlation functions as ($P_\mu O\equiv[P_\mu,O]=i\partial_\mu O$ from the algebra)
\begingroup\makeatletter\def\f@size{10}\check@mathfonts\def\maketag@@@#1{\hbox{\m@th\large\normalfont#1}}%
\eqna{
\langle Q^2O_i(x_1)\bar{Q}^2O_j^\dagger(x_2)[Q^2\bar{Q}^2O_{k(0,0)}]_p^\dagger(x_3)\rangle&=2^8\langle\mathcal{O}_i(z_1)\mathcal{O}_j^\dagger(z_2)\mathcal{O}_{k(0,0)}^\dagger(z_3)\rangle|_{\theta_1^2\bar{\theta}_2^2\theta_3^2\bar{\theta}_3^2}\\
&\phantom{=}\hspace{0.5cm}+2^6c_9P_3^{\dot{\gamma}\gamma}\langle\mathcal{O}_i(z_1)\mathcal{O}_j^\dagger(z_2)\mathcal{O}_{k(0,0)}^\dagger(z_3)\rangle|_{\theta_1^2\bar{\theta}_2^2\theta_3^\gamma\bar{\theta}_3^{\dot{\gamma}}}\\
&\phantom{=}\hspace{1cm}+2^7(c_{13}-c_1c_9)P_3^2\langle\mathcal{O}_i(z_1)\mathcal{O}_j^\dagger(z_2)\mathcal{O}_{k(0,0)}^\dagger(z_3)\rangle|_{\theta_1^2\bar{\theta}_2^2},\\
\langle O_i(x_1)O_j^\dagger(x_2)[Q^\xi\bar{Q}^{\dot{\xi}}O_{k(1,1)\xi\dot{\xi}}]_p^\dagger(x_3)\rangle&=-2^2\langle\mathcal{O}_i(z_1)\mathcal{O}_j^\dagger(z_2)(\mathcal{O}_{k(1,1)\beta\dot{\beta}})^\dagger(z_3)\rangle|_{\theta_{3\beta}\bar{\theta}_{3\dot{\beta}}}\\
&\phantom{=}\hspace{0.5cm}-c_4P_3^{\dot{\beta}\beta}\langle\mathcal{O}_i(z_1)\mathcal{O}_j^\dagger(z_2)(\mathcal{O}_{k(1,1)\beta\dot{\beta}})^\dagger(z_3)\rangle|,
}[EqnExample]
\endgroup
\textit{i.e.} as sums of specific components of the original three-point correlation function in terms of superfields.
\begin{table}[t]
\centering
\begin{tabular}{|c|c|}
\hline
$\langle\mathcal{O}_i\mathcal{O}_j(\mathcal{O}_{k(1,1)\beta\dot{\beta}})^\dagger\rangle$ & $\langle O_iO_j[Q^\xi\bar{Q}^{\dot{\xi}}O_{k(1,1)\xi\dot{\xi}}]_p^\dagger\rangle$\\\hline
$\varnothing$ & $-c_4P_3^{\dot{\beta}\beta}$\\
$\theta_{3\beta}\bar{\theta}_{3\dot{\beta}}$ & $-2^2$\\
\hline
\end{tabular}
\caption{Relevant linear combination for the $O_iO_j$ OPE with spin $1$ superfields $\mathcal{O}_{k(1,1)}$.  For the corresponding $Q^\alpha O_iQ_\alpha O_j$ OPE, the linear combinations consist of the same $\langle\mathcal{O}_i\mathcal{O}_j\mathcal{O}_{k(1,1)}^\dagger\rangle$ components with extra $\theta_1\theta_2$ and all coefficients must be multiplied by $2$.  For the corresponding $O_iO_j^\dagger$ OPE, the linear combinations are exactly the same with the same coefficients.  Finally, for the corresponding $Q^2O_i\bar{Q}^2O_j^\dagger$ OPE, the linear combinations consist of the same $\langle\mathcal{O}_i\mathcal{O}_j^\dagger\mathcal{O}_{k(1,1)}^\dagger\rangle$ components with extra $\theta_1^2\bar{\theta}_2^2$ and all coefficients must be multiplied by $2^4$.}
\label{TabOOO1}
\end{table}
\begin{table}[t]
\centering
\begin{tabular}{|c|c|}
\hline
$\langle\mathcal{O}_i\mathcal{O}_j(\mathcal{O}_{k(1,1)\beta\dot{\beta}})^\dagger\rangle$ & $\langle Q_\alpha O_iO_j[\bar{Q}^{\dot{\xi}}O_{k(1,1)\beta\dot{\xi}}]_p^\dagger\rangle$\\\hline
$\theta_1^\alpha\theta_{3\beta}$ & $2$\\
\hline
\end{tabular}
\caption{Relevant linear combination for the $Q_\alpha O_iO_j$ OPE with spin $1$ superfields $\mathcal{O}_{k(1,1)}$.  The quasi-primary field $[Q^2\bar{Q}^{\dot{\xi}}O_{k(1,1)\beta\dot{\xi}}]_p$ is not included here since its contribution vanishes once the remaining $Q^\alpha$ needed in the two-point correlation functions is taken into account.}
\label{TabQOOO1}
\end{table}
Therefore, the explicit form of the three-point correlation functions of superfields \eqref{EqnmOimOj} will be enough to retrieve the three-point correlation functions of fields.  Then the explicit form of the two-point correlation functions of superfields $\langle\mathcal{O}_{(j,\bar{\textit{\j}})}\mathcal{O}_{(j,\bar{\textit{\j}})}^\dagger\rangle$ will allow the computations of the two-point correlation functions of quasi-primary component fields.  Since this last step has already been done \cite{Li:2014gpa}, the focus next will primarily be on the three-point correlation functions of superfields \eqref{EqnmOimOj}.  With the two- and three-point correlation functions, it will then be straightforward to compute the OPE coefficients following \eqref{EqnOPEC}.

\subsection{Two- and Three-point Correlation Functions of Superfields}

The explicit form of the two- and three-point correlation functions of superfields is \cite{Osborn:1998qu}
\eqna{
\langle\mathcal{O}_i(z_1)\mathcal{O}_j^\dagger(z_2)\rangle&=\mathcal{C}_i\frac{I_{i\bar{\textit{\i}}}(x_{\bar{1}2},x_{\bar{2}1})}{x_{\bar{2}1}^{2q_i}x_{\bar{1}2}^{2\bar{q}_i}}\delta_{ij},\\
\langle\mathcal{O}_i(z_1)\mathcal{O}_j(z_2)\mathcal{O}_k^\dagger(z_3)\rangle&=\mathcal{C}_{ij\bar{k}}\frac{I_{i\bar{\textit{\i}}}(x_{\bar{1}3},x_{\bar{3}1})I_{j\bar{\textit{\j}}}(x_{\bar{2}3},x_{\bar{3}2})}{x_{\bar{3}1}^{2q_i}x_{\bar{1}3}^{2\bar{q}_i}x_{\bar{3}2}^{2q_j}x_{\bar{2}3}^{2\bar{q}_j}}t^{\bar{\textit{\i}}\bar{\textit{\j}}}_{\phantom{{\bar{\textit{\i}}\bar{\textit{\j}}}}k}(\bar{X}_3,\Theta_3,\bar{\Theta}_3),
}[EqnSCF]
where
\eqna{
x_{\bar{\textit{\i}}j}&=x_{ij}-i\theta_i\sigma\bar{\theta}_i-i\theta_j\sigma\bar{\theta}_j+2i\theta_j\sigma\bar{\theta}_i=-x_{j\bar{\textit{\i}}},\\
\bar{X}_3^\mu&=\frac{1}{2}\frac{x_{3\bar{2}\nu}x_{\bar{2}1\rho}x_{1\bar{3}\sigma}}{x_{\bar{2}3}^2x_{\bar{3}1}^2}\text{tr}(\bar{\sigma}^\mu\sigma^\nu\bar{\sigma}^\rho\sigma^\sigma)=(X_3^\mu)^\dagger,\\
\bar{\Theta}_3&=i\left(\frac{1}{x_{\bar{3}1}^2}\theta_{31}x_{1\bar{3}}\cdot\sigma-\frac{1}{x_{\bar{3}2}^2}\theta_{32}x_{2\bar{3}}\cdot\sigma\right)=\Theta_3^\dagger,
}
with $x_{ij}=x_i-x_j$, $\theta_{ij}=\theta_i-\theta_j$ and
\eqn{\bar{X}_3^2=\frac{x_{\bar{2}1}^2}{x_{\bar{2}3}^2x_{\bar{3}1}^2}.}
Here $\mathcal{C}_i$ and $\mathcal{C}_{ij\bar{k}}$ are the SUSic two- and three-point correlation function coefficients.  Moreover, the quantities $t^{\bar{\textit{\i}}\bar{\textit{\j}}}_{\phantom{{\bar{\textit{\i}}\bar{\textit{\j}}}}k}(\bar{X}_3,\Theta_3,\bar{\Theta}_3)$ make the three-point correlation functions of superfields transform coherently under the Lorentz group and have the following homogeneity property,
\eqn{t^{\bar{\textit{\i}}\bar{\textit{\j}}}_{\phantom{{\bar{\textit{\i}}\bar{\textit{\j}}}}k}(\lambda\bar{\lambda}\bar{X}_3,\lambda\Theta_3,\bar{\lambda}\bar{\Theta}_3)=\lambda^{\frac{2}{3}[2q_k+\bar{q}_k-(2q_i+\bar{q}_i+2q_j+\bar{q}_j)]}\bar{\lambda}^{\frac{2}{3}[q_k+2\bar{q}_k-(q_i+2\bar{q}_i+q_j+2\bar{q}_j)]}t^{\bar{\textit{\i}}\bar{\textit{\j}}}_{\phantom{{\bar{\textit{\i}}\bar{\textit{\j}}}}k}(\bar{X}_3,\Theta_3,\bar{\Theta}_3).}
They must also satisfy appropriate chirality properties.  For example, for chiral-chiral superfields $\mathcal{O}_i\mathcal{O}_j$ the quantity $t^{\bar{\textit{\i}}\bar{\textit{\j}}}_{\phantom{{\bar{\textit{\i}}\bar{\textit{\j}}}}k}(\bar{X}_3,\bar{\Theta}_3)$ is only a function of $\bar{X}_3$ and $\bar{\Theta}_3$ while for chiral-antichiral superfields $\mathcal{O}_i\mathcal{O}_j^\dagger$ the quantity $t^{\bar{\textit{\i}}\bar{\textit{\j}}}_{\phantom{{\bar{\textit{\i}}\bar{\textit{\j}}}}k}(\bar{X}_3)$ is only a function of $\bar{X}_3$.

The quantities $t^{\bar{\textit{\i}}\bar{\textit{\j}}}_{\phantom{{\bar{\textit{\i}}\bar{\textit{\j}}}}k}(\bar{X}_3,\Theta_3,\bar{\Theta}_3)$ needed for the relevant chiral-chiral and chiral-antichiral superfields $\mathcal{O}_i\mathcal{O}_j$ are already known in the literature \cite{Poland:2010wg}.

\subsubsection{Chiral-chiral Three-point Correlation Functions of Superfields}

For chiral-chiral superfields $\mathcal{O}_i\mathcal{O}_j$, \eqref{EqnSCF} simplifies to
\eqn{\langle\mathcal{O}_i(z_1)\mathcal{O}_j(z_2)\mathcal{O}_k^\dagger(z_3)\rangle=\mathcal{C}_{ij\bar{k}}\frac{1}{x_{\bar{3}1}^{2\Delta_i}x_{\bar{3}2}^{2\Delta_j}}t^{\bar{\textit{\i}}\bar{\textit{\j}}}_{\phantom{{\bar{\textit{\i}}\bar{\textit{\j}}}}k}(\bar{X}_3,\bar{\Theta}_3),}
and the homogeneity property implies that there are three different families of solutions given by \cite{Poland:2010wg}
\begin{itemize}
\item Solution I: Chiral operator ``$\mathcal{O}_i\mathcal{O}_j$'' with $(j_k,\bar{\textit{\j}}_k)=(0,0)$, $\Delta_k=\Delta_i+\Delta_j$ and $R_k=R_i+R_j$,\\
\eqn{t^{\bar{\textit{\i}}\bar{\textit{\j}}}_{\phantom{{\bar{\textit{\i}}\bar{\textit{\j}}}}k}(\bar{X}_3,\bar{\Theta}_3)=1;}
\item Solution II: Short operators with $(j_k,\bar{\textit{\j}}_k)=(\ell+1,\ell)$, $\Delta_k=\Delta_i+\Delta_j+\ell+1/2$ and $R_k=R_i+R_j-1$,\\
\eqn{t^{\bar{\textit{\i}}\bar{\textit{\j}}}_{\phantom{{\bar{\textit{\i}}\bar{\textit{\j}}}}k}(\bar{X}_3,\bar{\Theta}_3)=\bar{\Theta}_3^{(\dot{\alpha}_0}\bar{X}_{3\alpha_1}^{\dot{\alpha}_1}\cdots\bar{X}_{3\alpha_\ell}^{\dot{\alpha}_\ell)};}
\item Solution III: Long operators with $(j_k,\bar{\textit{\j}}_k)=(\ell,\ell)$ and $R_k=R_i+R_j-2$,\\
\eqn{t^{\bar{\textit{\i}}\bar{\textit{\j}}}_{\phantom{{\bar{\textit{\i}}\bar{\textit{\j}}}}k}(\bar{X}_3,\bar{\Theta}_3)=\bar{\Theta}_3^2\bar{X}_3^{\Delta_k-\Delta_i-\Delta_j-\ell-1}\bar{X}_3^{(\mu_1}\cdots\bar{X}_3^{\mu_\ell)}-\text{traces}.}
\end{itemize}
All remaining quasi-primary superfields $\mathcal{O}_k$ are forbidden by the unitarity bounds.  From the explicit forms of the quantities $t^{\bar{\textit{\i}}\bar{\textit{\j}}}_{\phantom{{\bar{\textit{\i}}\bar{\textit{\j}}}}k}(\bar{X}_3,\bar{\Theta}_3)$, spin $0$ superfields occur for type I and type III ($\ell=0$) three-point correlation functions, spin $1/2$ superfields occur for type II ($\ell=0$) three-point correlation functions and spin $1$ superfields occur for type III ($\ell=1$) three-point correlation functions.  As mentioned before, all spin $1/2$ superfields are short operators.

\subsubsection{Chiral-antichiral Three-point Correlation Functions of Superfields}

For chiral-antichiral superfields $\mathcal{O}_i\mathcal{O}_j^\dagger$, \eqref{EqnSCF} simplifies to
\eqn{\langle\mathcal{O}_i(z_1)\mathcal{O}_j^\dagger(z_2)\mathcal{O}_k^\dagger(z_3)\rangle=\mathcal{C}_{i\bar{\textit{\j}}\bar{k}}\frac{1}{x_{\bar{3}1}^{2\Delta_i}x_{\bar{2}3}^{2\Delta_j}}t^{\bar{\textit{\i}}\bar{\textit{\j}}}_{\phantom{{\bar{\textit{\i}}\bar{\textit{\j}}}}k}(\bar{X}_3),}
and the homogeneity property implies only one family of solutions given by \cite{Poland:2010wg}
\begin{itemize}
\item Long operators with $(j_k,\bar{\textit{\j}}_k)=(\ell,\ell)$ and $R_k=R_i-R_j$,\\
\eqn{t^{\bar{\textit{\i}}\bar{\textit{\j}}}_{\phantom{{\bar{\textit{\i}}\bar{\textit{\j}}}}k}(\bar{X}_3)=\bar{X}_3^{\Delta_k-\Delta_i-\Delta_j-\ell}\bar{X}_3^{(\mu_1}\cdots\bar{X}_3^{\mu_\ell)}-\text{traces}.}
\end{itemize}
Hence, only the identity (when $\mathcal{O}_j=\mathcal{O}_i$), spin $0$ superfields ($\ell=0$) and spin $1$ superfields ($\ell=1$) occur in three-point correlation functions of interest involving chiral-antichiral superfields $\mathcal{O}_i\mathcal{O}_j^\dagger$.

\subsection{Operator-product-expansion Coefficients}

With the knowledge of the appropriate three-point correlation functions of superfields, it is finally possible to compute the corrections to the visible-sector parameters from the OPE \eqref{EqnOiOj}.  For instance, continuing with the example \eqref{EqnExample} which corresponds to chiral-antichiral three-point correlation functions, one has
\eqna{
\langle Q^2O_i(x_1)\bar{Q}^2O_j^\dagger(x_2)[Q^2\bar{Q}^2O_{k(0,0)}]_p^\dagger(x_3)\rangle&=\frac{\frac{\substack{2^6(\Delta_i-\Delta_j-\Delta_k)(\Delta_i-\Delta_j-\Delta_k+2)(\Delta_i-\Delta_j+\Delta_k)\\\times(\Delta_i-\Delta_j+\Delta_k-2)(\Delta_i+\Delta_j+\Delta_k-2)(\Delta_i+\Delta_j+\Delta_k-4)}}{(\Delta_k-1)(\Delta_k-2)}\mathcal{C}_{ij\bar{k}}^{(0,0)}}{x_{12}^{\Delta_i+\Delta_j-\Delta_k}x_{13}^{\Delta_i-\Delta_j+\Delta_k+2}x_{23}^{-\Delta_i+\Delta_j+\Delta_k+2}},\\
\langle O_i(x_1)O_j^\dagger(x_2)[Q^\xi\bar{Q}^{\dot{\xi}}O_{k(1,1)\xi\dot{\xi}}]_p^\dagger(x_3)\rangle&=\frac{-\frac{2i(\Delta_i-\Delta_j-\Delta_k+3)(\Delta_i-\Delta_j+\Delta_k-3)}{(\Delta_k-3)}\mathcal{C}_{ij\bar{k}}^{(1,1)}}{x_{12}^{\Delta_i+\Delta_j-\Delta_k-1}x_{13}^{\Delta_i-\Delta_j+\Delta_k+1}x_{23}^{-\Delta_i+\Delta_j+\Delta_k+1}}.
}
Since from \eqref{EqnSCF} the relevant two-point correlation functions are given by \cite{Li:2014gpa}
\eqna{
\langle[Q^2\bar{Q}^2O_{k(0,0)}]_p(x)[Q^2\bar{Q}^2O_{k(0,0)}]_p^\dagger(0)\rangle&=\frac{\frac{\substack{2^8\Delta_k(\Delta_k+1)(\Delta_i-\Delta_j-\Delta_k)(\Delta_i-\Delta_j-\Delta_k+2)\\\times(\Delta_i-\Delta_j+\Delta_k)(\Delta_i-\Delta_j+\Delta_k-2)}}{(\Delta_k-1)(\Delta_k-2)}\mathcal{C}_k^{(0,0)}}{x^{2(\Delta_k+2)}},\\
\langle[Q^\gamma\bar{Q}^{\dot{\gamma}}O_{k(1,1)\gamma\dot{\gamma}}]_p(x)[Q^\xi\bar{Q}^{\dot{\xi}}O_{k(1,1)\xi\dot{\xi}}]_p^\dagger(0)\rangle&=\frac{\frac{2^4(\Delta_k-2)(\Delta_i-\Delta_j-\Delta_k+3)(\Delta_i-\Delta_j+\Delta_k-3)}{(\Delta_k-3)}\mathcal{C}_k^{(1,1)}}{x^{2(\Delta_k+1)}},
}
the OPE coefficients are straightforwardly obtained from \eqref{EqnOPEC} in the limit $x_{12}\to0$ as
\eqna{
\lambda_{Q^2O_i\bar{Q}^2O_j^\dagger}^{\phantom{Q^2O_i\bar{Q}^2O_j^\dagger}Q^2\bar{Q}^2O_{k(0,0)}}&=\frac{(\Delta_i+\Delta_j+\Delta_k-2)(\Delta_i+\Delta_j+\Delta_k-4)}{2^2\Delta_k(\Delta_k+1)}\frac{\mathcal{C}_{ij\bar{k}}^{(0,0)}}{\mathcal{C}_k^{(0,0)}},\\
\lambda_{O_iO_j^\dagger}^{\phantom{O_iO_j^\dagger}Q\bar{Q}O_{k(1,1)}}&=-\frac{i}{2^3(\Delta_k-2)}\frac{\mathcal{C}_{ij\bar{k}}^{(1,1)}}{\mathcal{C}_k^{(1,1)}}.
}
Here the obvious normalization for the differential operators $\mathcal{D}_{ij}^{\phantom{ij}k}(x_2,\partial_2)$ has been chosen.  The remaining OPE coefficients are computed below.


\section{Results for the Operator Product Expansions}\label{SOPEResults}

In this section the relevant OPEs are computed with the help of the techniques described previously.  Although the computations are straightforward, they are tedious enough to justify the need for computers.  The results shown here were obtained with the help of \textit{Mathematica}.

In the following OPEs, the ellipses stand for contributions to the OPEs that vanish for the two-point correlation functions \eqref{EqnBraOiOjKet}.  For instance, the ellipses in $(1+\cdots)$ stand for derivatives from the normalized differential operators $\mathcal{D}_{ij}^{\phantom{ij}k}(x,-iP)$ and the ellipses at the end include all contributions with vanishing vevs.  Furthermore, conformal dimensions always correspond to those of the superfields, not the associated quasi-primary component fields.

Moreover, the most singular term in the differential operators relevant to quasi-primary operators from type I and type II solutions do not depend on the separation $|x|$.  This observation will be important when computing the corrections to the visible-sector parameters.

\subsection{\texorpdfstring{$O_iO_j$}{O_iO_j} Operator Product Expansions}

The only non-trivial contributions to the $O_iO_j$ OPEs come from type I and spin $0$ type III solutions.  They are given by
\begingroup\makeatletter\def\f@size{10}\check@mathfonts\def\maketag@@@#1{\hbox{\m@th\large\normalfont#1}}%
\eqn{O_i(x)O_j(0)=\sum_k\left[\lambda_{O_iO_j}^{\phantom{O_iO_j}O_{k(0,0)}^\text{I}}(1+\cdots)O_{k(0,0)}^\text{I}(0)+\lambda_{O_iO_j}^{\phantom{O_iO_j}\bar{Q}^2O_{k(0,0)}^\text{III}}\frac{(1+\cdots)}{x^{\Delta_i+\Delta_j-\Delta_k-1}}\bar{Q}^2O_{k(0,0)}^\text{III}(0)\right]+\cdots,}[EqnOO]
\endgroup
where the OPE coefficients are
\eqn{
\begin{gathered}
\lambda_{O_iO_j}^{\phantom{O_iO_j}O_{k(0,0)}^\text{I}}=\frac{\mathcal{C}_{ij\bar{k}}^{\text{I}(0,0)}}{\mathcal{C}_k^{\text{I}(0,0)}},\quad\quad\lambda_{O_iO_j}^{\phantom{O_iO_j}\bar{Q}^2O_{k(0,0)}^\text{III}}=-\frac{1}{2^2(\Delta_i+\Delta_j-\Delta_k-1)(\Delta_i+\Delta_j-\Delta_k-3)}\frac{\mathcal{C}_{ij\bar{k}}^{\text{III}(0,0)}}{\mathcal{C}_k^{\text{III}(0,0)}}.
\end{gathered}
}[EqnOOC]
Here it is important to observe that only quasi-primary operators from type I solutions appear in the two-point correlation functions obtained from \eqref{EqnOO} when SUSY is conserved.  These terms will be shown to disappear in the corrections to the visible-sector parameters.

\subsection{\texorpdfstring{$Q_\alpha O_iO_j$}{Q_\alpha O_iO_j} Operator Product Expansions}

For the $Q_\alpha O_iO_j$ OPEs, there are non-trivial contributions from all three types of solutions.  Moreover, for type III both spin $0$ and spin $1$ solutions occur.  The OPE is
\begingroup\makeatletter\def\f@size{10}\check@mathfonts\def\maketag@@@#1{\hbox{\m@th\large\normalfont#1}}%
\eqna{
Q_\alpha O_i(x)O_j(0)&=\sum_k\left[\lambda_{QO_iO_j}^{\phantom{QO_iO_j}QO_{k(0,0)}^\text{I}}(1+\cdots)Q_\alpha O_{k(0,0)}^\text{I}(0)+\lambda_{QO_iO_j}^{\phantom{QO_iO_j}O_{k(1,0)}^\text{II}}(1+\cdots)O_{k(1,0)\alpha}^\text{II}(0)\right.\\
&\phantom{=}\hspace{0.5cm}\left.+\lambda_{QO_iO_j}^{\phantom{QO_iO_j}\bar{Q}^2QO_{k(0,0)}^\text{III}}\frac{(1+\cdots)}{x^{\Delta_i+\Delta_j-\Delta_k-1}}[\bar{Q}^2Q_\alpha O_{k(0,0)}^\text{III}]_p(0)\right.\\
&\phantom{=}\hspace{1cm}\left.+\lambda_{QO_iO_j}^{\phantom{QO_iO_j}\bar{Q}O_{k(1,1)}^\text{III}}\frac{(1+\cdots)}{x^{\Delta_i+\Delta_j-\Delta_k}}\bar{Q}^{\dot{\alpha}}O_{k(1,1)\alpha\dot{\alpha}}^\text{III}(0)\right]+\cdots,
}[EqnQOO]
\endgroup
and the OPE coefficients are
\begingroup\makeatletter\def\f@size{10}\check@mathfonts\def\maketag@@@#1{\hbox{\m@th\large\normalfont#1}}%
\eqn{
\begin{gathered}
\lambda_{QO_iO_j}^{\phantom{QO_iO_j}QO_{k(0,0)}^\text{I}}=\frac{\Delta_i}{\Delta_i+\Delta_j}\frac{\mathcal{C}_{ij\bar{k}}^{\text{I}(0,0)}}{\mathcal{C}_k^{\text{I}(0,0)}},\quad\quad\lambda_{QO_iO_j}^{\phantom{QO_iO_j}O_{k(1,0)}^\text{II}}=-\frac{\mathcal{C}_{ij\bar{k}}^{\text{II}(1,0)}}{\mathcal{C}_k^{\text{II}(1,0)}},\\
\lambda_{QO_iO_j}^{\phantom{QO_iO_j}\bar{Q}^2QO_{k(0,0)}^\text{III}}=-\frac{(\Delta_k-1)(\Delta_i-\Delta_j+\Delta_k+1)}{2^3(\Delta_k+1)(\Delta_i+\Delta_j-\Delta_k-1)(\Delta_i+\Delta_j-\Delta_k-3)(\Delta_i+\Delta_j+\Delta_k-3)}\frac{\mathcal{C}_{ij\bar{k}}^{\text{III}(0,0)}}{\mathcal{C}_k^{\text{III}(0,0)}},\\
\lambda_{QO_iO_j}^{\phantom{QO_iO_j}\bar{Q}O_{k(1,1)}^\text{III}}=\frac{i}{2(\Delta_i+\Delta_j-\Delta_k)}\frac{\mathcal{C}_{ij\bar{k}}^{\text{III}(1,1)}}{\mathcal{C}_k^{\text{III}(1,1)}}.
\end{gathered}
}[EqnQOOC]
\endgroup
It is clear the two-point correlation functions of \eqref{EqnQOO} with an extra charge $Q^\alpha$ vanish in the SUSic limit.

\subsection{\texorpdfstring{$Q^\alpha O_iQ_\alpha O_j$}{Q^\alpha O_iQ_\alpha O_j} Operator Product Expansions}

There are five non-vanishing contributions to the $Q^\alpha O_iQ_\alpha O_j$ OPEs coming from all three types of solutions and choices of spins.  In particular, there are two quasi-primary component fields appearing from spin $0$ type III solutions.  The OPE is expressed as
\begingroup\makeatletter\def\f@size{10}\check@mathfonts\def\maketag@@@#1{\hbox{\m@th\large\normalfont#1}}%
\eqna{
Q^\alpha O_i(x)Q_\alpha O_j(0)&=\sum_k\left[\lambda_{QO_iQO_j}^{\phantom{QO_iQO_j}Q^2O_{k(0,0)}^\text{I}}(1+\cdots)Q^2O_{k(0,0)}^\text{I}(0)+\lambda_{QO_iQO_j}^{\phantom{QO_iQO_j}QO_{k(1,0)}^\text{II}}(1+\cdots)Q^\alpha O_{k(1,0)\alpha}^\text{II}(0)\right.\\
&\phantom{=}\hspace{0.5cm}\left.+\lambda_{QO_iQO_j}^{\phantom{QO_iQO_j}O_{k(0,0)}^\text{III}}\frac{(1+\cdots)}{x^{\Delta_i+\Delta_j-\Delta_k+1}}O_{k(0,0)}^\text{III}(0)\right.\\
&\phantom{=}\hspace{1cm}\left.+\lambda_{QO_iQO_j}^{\phantom{QO_iQO_j}Q^2\bar{Q}^2O_{k(0,0)}^\text{III}}\frac{(1+\cdots)}{x^{\Delta_i+\Delta_j-\Delta_k-1}}[Q^2\bar{Q}^2O_{k(0,0)}^\text{III}]_p(0)\right.\\
&\phantom{=}\hspace{1.5cm}\left.+\lambda_{QO_iQO_j}^{\phantom{QO_iQO_j}Q\bar{Q}O_{k(1,1)}^\text{III}}\frac{(1+\cdots)}{x^{\Delta_i+\Delta_j-\Delta_k}}[Q^\alpha\bar{Q}^{\dot{\alpha}}O_{k(1,1)\alpha\dot{\alpha}}^\text{III}]_p(0)\right]+\cdots,
}[EqnQOQO]
\endgroup
while the OPE coefficients are
\eqn{
\begin{gathered}
\lambda_{QO_iQO_j}^{\phantom{QO_iQO_j}Q^2O_{k(0,0)}^\text{I}}=\frac{\Delta_i\Delta_j}{(\Delta_i+\Delta_j)(\Delta_i+\Delta_j-1)}\frac{\mathcal{C}_{ij\bar{k}}^{\text{I}(0,0)}}{\mathcal{C}_k^{\text{I}(0,0)}},\quad\quad\lambda_{QO_iQO_j}^{\phantom{QO_iO_j}QO_{k(1,0)}^\text{II}}=\frac{\Delta_i-\Delta_j}{\Delta_i+\Delta_j-2}\frac{\mathcal{C}_{ij\bar{k}}^{\text{II}(1,0)}}{\mathcal{C}_k^{\text{II}(1,0)}},\\
\lambda_{QO_iQO_j}^{\phantom{QO_iQO_j}O_{k(0,0)}^\text{III}}=2^2\frac{\mathcal{C}_{ij\bar{k}}^{\text{III}(0,0)}}{\mathcal{C}_k^{\text{III}(0,0)}},\\
\lambda_{QO_iQO_j}^{\phantom{QO_iQO_j}Q^2\bar{Q}^2O_{k(0,0)}^\text{III}}=\frac{(\Delta_i-\Delta_j-\Delta_k-1)(\Delta_i-\Delta_j+\Delta_k+1)}{2^4\Delta_k(\Delta_k+1)(\Delta_i+\Delta_j-\Delta_k-1)(\Delta_i+\Delta_j-\Delta_k-3)}\frac{\mathcal{C}_{ij\bar{k}}^{\text{III}(0,0)}}{\mathcal{C}_k^{\text{III}(0,0)}},\\
\lambda_{QO_iQO_j}^{\phantom{QO_iQO_j}Q\bar{Q}O_{k(1,1)}^\text{III}}=-\frac{i(\Delta_i-\Delta_j)}{2(\Delta_k-2)(\Delta_i+\Delta_j-\Delta_k)}\frac{\mathcal{C}_{ij\bar{k}}^{\text{III}(1,1)}}{\mathcal{C}_k^{\text{III}(1,1)}}.
\end{gathered}
}[EqnQOQOC]
In the SUSic limit, only contributions from spin $0$ type III quasi-primary operators occur in the two-point correlation functions of \eqref{EqnQOQO}.  These contributions modify the parameters of the superpotential even when SUSY is unbroken and correspond to the usual modifications expected when integrating out SUSic degrees of freedom.

\subsection{\texorpdfstring{$O_iO_j^\dagger$}{O_iO_j^\dagger} Operator Product Expansions}

For the $O_iO_j^\dagger$ OPEs, there are one contribution from the identity, two non-trivial contributions from spin $0$ superfields, and one non-trivial contribution from spin $1$ superfields,
\begingroup\makeatletter\def\f@size{10}\check@mathfonts\def\maketag@@@#1{\hbox{\m@th\large\normalfont#1}}%
\eqna{
O_i(x)O_j^\dagger(0)&=\mathcal{C}_i\frac{\delta_{ij}}{x^{2\Delta_i}}\mathds{1}+\sum_k\left[\lambda_{O_iO_j^\dagger}^{\phantom{O_iO_j^\dagger}O_{k(0,0)}}\frac{(1+\cdots)}{x^{\Delta_i+\Delta_j-\Delta_k}}O_{k(0,0)}(0)\right.\\
&\phantom{=}\hspace{0.5cm}\left.+\lambda_{O_iO_j^\dagger}^{\phantom{O_iO_j^\dagger}Q^2\bar{Q}^2O_{k(0,0)}}\frac{(1+\cdots)}{x^{\Delta_i+\Delta_j-\Delta_k-2}}[Q^2\bar{Q}^2O_{k(0,0)}]_p(0)\right.\\
&\phantom{=}\hspace{1cm}\left.+\lambda_{O_iO_j^\dagger}^{\phantom{O_iO_j^\dagger}Q\bar{Q}O_{k(1,1)}}\frac{(1+\cdots)}{x^{\Delta_i+\Delta_j-\Delta_k-1}}[Q^\alpha\bar{Q}^{\dot{\alpha}}O_{k(1,1)\alpha\dot{\alpha}}]_p(0)\right]+\cdots.
}[EqnOOdag]
\endgroup
The OPE coefficients can be written as
\eqn{
\begin{gathered}
\lambda_{O_iO_j^\dagger}^{\phantom{O_iO_j^\dagger}O_{k(0,0)}}=\frac{\mathcal{C}_{i\bar{\textit{\j}}\bar{k}}^{(0,0)}}{\mathcal{C}_k^{(0,0)}},\quad\quad\lambda_{O_iO_j^\dagger}^{\phantom{O_iO_j^\dagger}Q^2\bar{Q}^2O_{k(0,0)}}=\frac{1}{2^6\Delta_k(\Delta_k+1)}\frac{\mathcal{C}_{i\bar{\textit{\j}}\bar{k}}^{(0,0)}}{\mathcal{C}_k^{(0,0)}},\\
\lambda_{O_iO_j^\dagger}^{\phantom{O_iO_j^\dagger}Q\bar{Q}O_{k(1,1)}}=-\frac{i}{2^3(\Delta_k-2)}\frac{\mathcal{C}_{i\bar{\textit{\j}}\bar{k}}^{(1,1)}}{\mathcal{C}_k^{(1,1)}}.
\end{gathered}
}[EqnOOdagC]
The identity operator and spin $0$ operators can generate non-trivial $\langle O_iO_j^\dagger\rangle$ two-point correlation functions in the SUSic limit.\footnote{Note that the identity operator can be grouped with the spin $0$ operators.}  Those contributions lead to SUSic wave-function renormalizations, as expected.

\subsection{\texorpdfstring{$Q^2O_i\bar{Q}^2O_j^\dagger$}{Q^2O_i\bar{Q}^2O_j^\dagger} Operator Product Expansions}

For the OPEs of the products of $Q^2O_i$ and $\bar{Q}^2O_j^\dagger$, there are also one contribution from the identity, two non-trivial contributions from spin $0$ superfields, and one non-trivial contribution from spin $1$ superfields, leading to
\begingroup\makeatletter\def\f@size{10}\check@mathfonts\def\maketag@@@#1{\hbox{\m@th\large\normalfont#1}}%
\eqna{
Q^2O_i(x)\bar{Q}^2O_j^\dagger(0)&=2^6\Delta_i(\Delta_i-1)\mathcal{C}_i\frac{\delta_{ij}}{x^{2(\Delta_i+1)}}\mathds{1}+\sum_k\left[\lambda_{Q^2O_i\bar{Q}^2O_j^\dagger}^{\phantom{Q^2O_i\bar{Q}^2O_j^\dagger}O_{k(0,0)}}\frac{(1+\cdots)}{x^{\Delta_i+\Delta_j-\Delta_k+2}}O_{k(0,0)}(0)\right.\\
&\phantom{=}\hspace{0.5cm}\left.+\lambda_{Q^2O_i\bar{Q}^2O_j^\dagger}^{\phantom{Q^2O_i\bar{Q}^2O_j^\dagger}Q^2\bar{Q}^2O_{k(0,0)}}\frac{(1+\cdots)}{x^{\Delta_i+\Delta_j-\Delta_k}}[Q^2\bar{Q}^2O_{k(0,0)}]_p(0)\right.\\
&\phantom{=}\hspace{1cm}\left.+\lambda_{Q^2O_i\bar{Q}^2O_j^\dagger}^{\phantom{Q^2O_i\bar{Q}^2O_j^\dagger}Q\bar{Q}O_{k(1,1)}}\frac{(1+\cdots)}{x^{\Delta_i+\Delta_j-\Delta_k+1}}[Q^\alpha\bar{Q}^{\dot{\alpha}}O_{k(1,1)\alpha\dot{\alpha}}]_p(0)\right]+\cdots.
}[EqnQsqOQbsqOdag]
\endgroup
The coefficients of the identity operator can be found easily from the two-point correlation functions (see \cite{Li:2014gpa}).  They are simply related to the coefficients of the identity operators in \eqref{EqnOOdag}.  The remaining OPE coefficients can be expressed as
\eqna{
\lambda_{Q^2O_i\bar{Q}^2O_j^\dagger}^{\phantom{Q^2O_i\bar{Q}^2O_j^\dagger}O_{k(0,0)}}&=2^4(\Delta_i+\Delta_j-\Delta_k)(\Delta_i+\Delta_j-\Delta_k-2)\frac{\mathcal{C}_{i\bar{\textit{\j}}\bar{k}}^{(0,0)}}{\mathcal{C}_k^{(0,0)}},\\
\lambda_{Q^2O_i\bar{Q}^2O_j^\dagger}^{\phantom{Q^2O_i\bar{Q}^2O_j^\dagger}Q^2\bar{Q}^2O_{k(0,0)}}&=\frac{(\Delta_i+\Delta_j+\Delta_k-2)(\Delta_i+\Delta_j+\Delta_k-4)}{2^2\Delta_k(\Delta_k+1)}\frac{\mathcal{C}_{i\bar{\textit{\j}}\bar{k}}^{(0,0)}}{\mathcal{C}_k^{(0,0)}},\\
\lambda_{Q^2O_i\bar{Q}^2O_j^\dagger}^{\phantom{Q^2O_i\bar{Q}^2O_j^\dagger}Q\bar{Q}O_{k(1,1)}}&=-\frac{2i(\Delta_i+\Delta_j-\Delta_k-3)(\Delta_i+\Delta_j+\Delta_k-5)}{\Delta_k-2}\frac{\mathcal{C}_{i\bar{\textit{\j}}\bar{k}}^{(1,1)}}{\mathcal{C}_k^{(1,1)}}.
}[EqnQsqOQbsqOdagC]
For unbroken SUSY, the identity operator and spin $0$ operators can again give non-trivial contributions to the OPEs.\footnote{Again note that the identity operator can be grouped with spin $0$ operators.}  Hence, this suggests that the vanishing of $\langle Q^2O_i\bar{Q}^2O_j^\dagger\rangle$ at zero momentum in the SUSic limit implies non-trivial relations between the vevs of the operators of interest, \textit{i.e.} the identity operator and spin $0$ operators $O_{k(0,0)}$.


\section{Visible-sector Corrections}\label{SResults}

This section derives the corrections to the visible-sector parameters from the OPEs and dispersion relations.  The dispersion relations explain why type I and type II solutions do not contribute to the visible-sector corrections.  Some general comments about the implications of the OPE techniques are also given.

\subsection{Dispersion Relations}

\begin{figure}[t!]
\centering
\resizebox{11cm}{!}{
\includegraphics{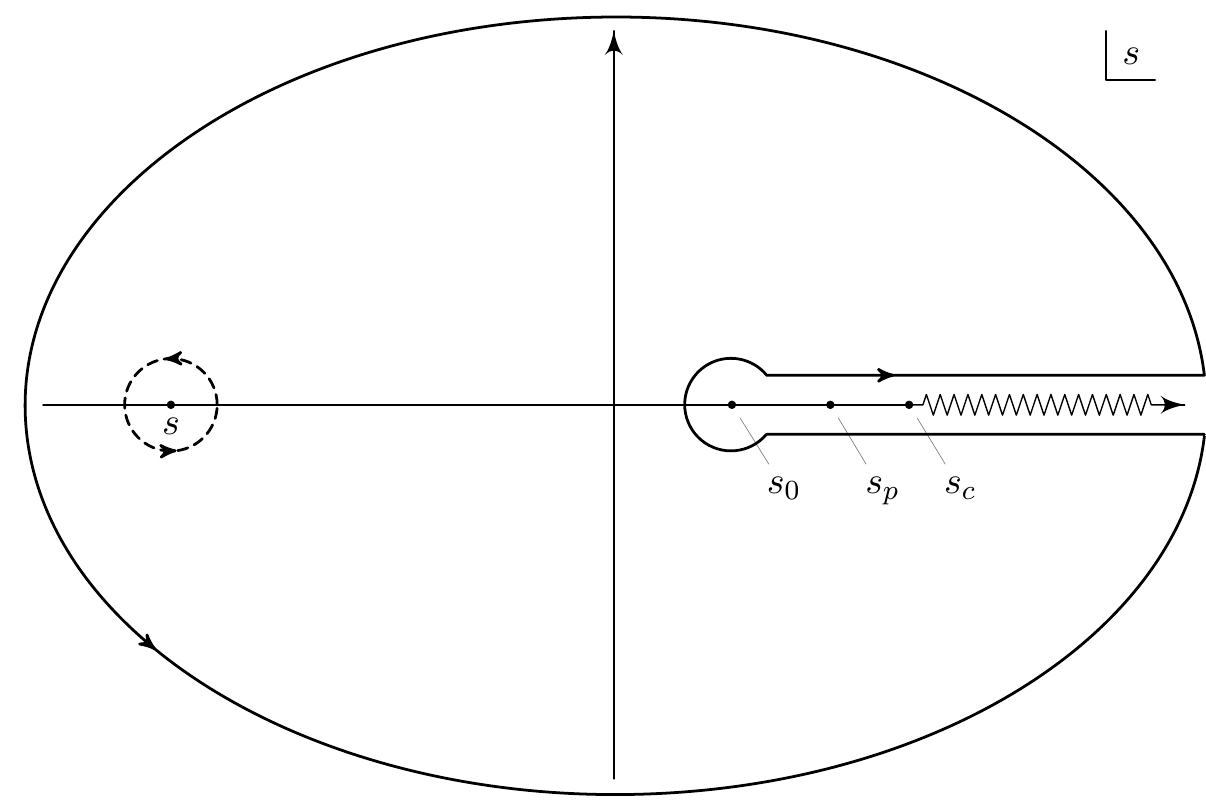}
}
\caption{Dispersion relation for amplitudes $\tilde{A}(s)$.  The dashed contour is deformed to the solid contour, leading to \eqref{EqnDispersionRelation}.}
\label{FigDispersionRelation}
\end{figure}

Since the OPE is a short-distance/large-momentum expansion, dispersion relations might be necessary to properly evaluate the corrections.  Using the usual dispersion relation\footnote{The discontinuity and the imaginary part should satisfy $\text{Disc}\,\tilde{A}(s')=2i\,\text{Im}\,\tilde{A}(s')$ as for two-to-two elastic scattering of spin $0$ particles.} pictured in figure~\ref{FigDispersionRelation}, \textit{i.e.}
\eqn{\tilde{A}(s)=\frac{1}{2\pi i}\int_{s_0}^\infty ds'\,\frac{\text{Disc}\,\tilde{A}(s')}{s'-s}=\frac{1}{\pi}\int_{s_0}^\infty ds'\,\frac{\text{Im}\,\tilde{A}(s')}{s'-s},}[EqnDispersionRelation]
for an amplitude $\tilde{A}(s)=i\int d^4x\,e^{-ip\cdot x}A(x)$ (where $s=-p^2$) with branch cuts starting at $s_{c_i}$ coming from multi-particle states and single poles $s_{p_i}$ coming from fundamental one-particle states or bound states on the physical sheet where the threshold $s_0$ satisfies $s_0<s_{c_i}$ and $s_0<s_{p_i}$, all the corrections to the visible-sector parameters can be computed easily.

For a two-point correlation function $A(x)$, the visible-sector corrections \eqref{EqndWV} to \eqref{EqndLVo} are given by two types of integrals:
\eqn{\int d^4x\,A(x),\quad\quad\text{or}\quad\quad\int d^4x\,\delta^{(4)}(x)\frac{1}{\partial^2}A(x).}
The dispersion relation \eqref{EqnDispersionRelation} is therefore needed for both integrals since the first type of integrals corresponds to zero-momentum Fourier transforms $\text{lim}_{p\to0}\int d^4x\,e^{-ip\cdot x}$ while the second type of integrals involves an integration over the negative $s$ axis,
\eqn{
\begin{gathered}
\int d^4x\,A(x)=-i\tilde{A}(0),\\
\int d^4x\,\delta^{(4)}(x)\frac{1}{\partial^2}A(x)=-\frac{1}{16\pi^2}\int_0^\infty ds\,\tilde{A}(-s).
\end{gathered}
}[EqnInt]
Hence both integrals use information away from the short-distance/large-momentum expansion provided by the OPE.

With the OPE, the visible-sector corrections relate simply to the particular case $A(x)=x^h$ for an appropriate power $h$.  Therefore, since for $x^h$ the Fourier transform is
\eqn{\int d^4x\,e^{-ip\cdot x}x^h=2^{h+3}\pi i(h+2)\Gamma^2\left(1+\frac{h}{2}\right)\sin\left(\frac{\pi h}{2}\right)\frac{1}{(-s)^{h/2+2}},}
by analytic continuation, the integrals \eqref{EqnInt} above become\footnote{The second integral in \eqref{EqnIntOPE} was regulated as in \cite{Fortin:2011ad}.}
\eqna{
\int d^4x\,x^h&=i\frac{h+2}{h+4}\Gamma^2\left(1+\frac{h}{2}\right)\sin^2\left(\frac{\pi h}{2}\right)\left(\frac{4}{s_0}\right)^{h/2+2},\\
\int d^4x\,\delta^{(4)}(x)\frac{1}{\partial^2}x^h&=-\frac{1}{2\pi^2}\frac{1}{h+2}\Gamma^2\left(1+\frac{h}{2}\right)\sin^2\left(\frac{\pi h}{2}\right)\left(\frac{4}{s_0}\right)^{h/2+1},
}[EqnIntOPE]
where, as suggested by the K\"{a}ll\'en-Lehmann spectral representation in the SUSic limit, the threshold for branch cuts only is taken to be at $s_0=4M^2$ while the threshold for single poles is taken to be at $s_0=M^2$ with $M$ the typical hidden-sector mass scale.

Although it has been explained carefully in \cite{Fortin:2011ad,Fortin:2012tp,Kumar:2014uxa}, it is important to enumerate explicitly all the approximations made in the computations of the corrections to the visible-sector parameters presented here.  First, the two-point correlation functions $A(x)$ are approximated by their OPEs, effectively taking the short-distance limit.  Second, the OPEs, which are valid in the short-distance/large-momentum limit, are nevertheless used in the entire region of integration from the thresholds to $\infty$.  Third, the thresholds and singularities are chosen to always be the SUSic thresholds $4M^2$ and $M^2$ respectively, even though SUSY-breaking effects change the position of the branch points and single poles by correcting the masses of hidden-sector fields.

Therefore, at quadratic order in the coupling constants, the corrections to the visible-sector parameters in terms of one- and two-point correlation functions of hidden fields \eqref{EqndWV}, \eqref{EqndLVs} and \eqref{EqndLVo} are exact.  Moreover, the OPEs \eqref{EqnOO}, \eqref{EqnQOO}, \eqref{EqnQOQO}, \eqref{EqnOOdag} and \eqref{EqnQsqOQbsqOdag} with the OPE coefficients \eqref{EqnOOC}, \eqref{EqnQOOC}, \eqref{EqnQOQOC}, \eqref{EqnOOdagC} and \eqref{EqnQsqOQbsqOdagC} are also exact.  However, in light of what is stated above, the results presented below are approximations, although they are known to give back the full answer in some special cases (see \cite{Fortin:2011ad}).

\subsection{Results and Comments}

The final results can be obtained by using the OPEs \eqref{EqnOO}, \eqref{EqnQOO}, \eqref{EqnQOQO}, \eqref{EqnOOdag} and \eqref{EqnQsqOQbsqOdag} with the OPE coefficients \eqref{EqnOOC}, \eqref{EqnQOOC}, \eqref{EqnQOQOC}, \eqref{EqnOOdagC} and \eqref{EqnQsqOQbsqOdagC} in the visible-sector corrections \eqref{EqndWV}, \eqref{EqndLVs} and \eqref{EqndLVo} where the integrals are given in \eqref{EqnIntOPE}.

Several comments are in order, first on the contributions to the visible-sector corrections and then on the phenomenology.  Indeed, the unified approach presented here with the general formalism of the OPE techniques lead to a very structured understanding of some properties of specific models, as well as possible solutions to generic phenomenological problems.

First, for the corrections \eqref{EqndWV}, \eqref{EqndLVs} and \eqref{EqndLVo}, since both integrals \eqref{EqnIntOPE} vanish for non-negative integers $h/2$, all contributions to the visible-sector corrections originating from type I and type II solutions vanish, explaining why $\lambda^{ab}$ vanish in the SUSic limit.  Hence, only long operators contribute to the corrections to the visible-sector parameters.\footnote{The divergences in \eqref{EqnIntOPE} at $h=-4$ and $h=-2$ respectively correspond to logarithmic divergences expected, for example, in wave-function renormalizations.}

Then, several general comments on the phenomenology can be made.  For example, the soft Lagrangian parameters $a^{abc}$ and $b^{ab}$ are $-\frac{1}{4}Q^2$ times the corrections to the superpotential $\delta y^{abc}$ and $\delta M^{ab}$ respectively.  Hence, the OPE coefficients for these soft Lagrangian parameters are simply $-\frac{1}{4}$ times the OPE coefficients of the corresponding superpotential corrections.  Therefore the different vevs of the long operators on the left of the OPEs, either with $Q^2$ for the soft Lagrangian parameters $a^{abc}$ and $b^{ab}$ or without $Q^2$ for the superpotential corrections $\delta y^{abc}$ and $\delta M^{ab}$, dictate the relative size of these parameters, not the OPE coefficients.  Since some operators on the left of the OPEs vanish when $Q^2$ acts on them (for example $[Q^2\bar{Q}^2O_{k(0,0)}^\text{III}]_p$), there are necessarily fewer contributions to these soft Lagrangian parameters.  As discussed in \cite{Kumar:2014uxa}, this has obvious implications for the $\mu/B_\mu$ problem \cite{Dvali:1996cu}, although in general the soft Lagrangian contributions to $B_\mu$ from $u^a_{\phantom{a}b}$ and $v_a$ must be considered.  Moreover, there are also clear implications for the $A/m_H^2$ problem \cite{Craig:2013wga} since the soft Lagrangian parameters $(m^2)^a_{\phantom{a}b}$ are independent from the trilinear couplings.  Although this is discussed in \cite{Kumar:2014uxa}, technically, their analysis does not use the proper OPE for the evaluation of $\delta m_{H_{u,d}}^2$.  However, as mentioned below \eqref{EqndLVo}, since the relevant integral for $\delta m_{H_{u,d}}^2$ is a zero-momentum Fourier transform, the results should be independent of the OPE chosen [either \eqref{EqnOOdag} or \eqref{EqnQsqOQbsqOdag}].  Indeed, in the specific framework of \cite{Kumar:2014uxa}, there are contributions to the trilinear couplings from $Q^2O_{k(0,0)}$ and there are contributions to $\delta m_H^2$ only from $[Q^2\bar{Q}^2O_{k(0,0)}]_p$.  The appropriate results shown here demonstrate that there are contributions to $\delta m_H^2$ from $O_{k(0,0)}$, $[Q^2\bar{Q}^2O_{k(0,0)}]_p$ and $[Q^\alpha\bar{Q}^{\dot{\alpha}}O_{k(1,1)\alpha\dot{\alpha}}]_p$ instead of only $[Q^2\bar{Q}^2O_{k(0,0)}]_p$ as stated in \cite{Kumar:2014uxa}.\footnote{The work of \cite{Kumar:2014uxa} could have been used directly for most OPEs of interest here.  The results presented here are however more complete.  Moreover, they differ slightly from the results of \cite{Kumar:2014uxa} in two other instances.  First, there is a typo for the $\lambda_{QO_iO_j}^{\phantom{QO_iO_j}O_{k(1,0)}^\text{II}}$ OPE coefficient and second, the $\lambda_{QO_iO_j}^{\phantom{QO_iO_j}\bar{Q}^2QO_{k(0,0)}^\text{III}}$ OPE coefficient is wrong.}  The leading contribution to $\delta m_H^2$ therefore comes from $O_{k(0,0)}$ operators, although on general grounds one must be careful about the expected cancellation mentioned below \eqref{EqnQsqOQbsqOdagC}.  In fact, in the weakly-coupled example of \cite{Kumar:2014uxa}, it was necessary to include all the OPE contributions to $\delta m_H^2$ to show that it vanishes.  It is thus expected that all the OPE contributions from $O_{k(0,0)}$ and $[Q^\alpha\bar{Q}^{\dot{\alpha}}O_{k(1,1)\alpha\dot{\alpha}}]_p$ must be included to explicitly show the equivalence between the two OPEs \textit{inside} a zero-momentum Fourier transform.


\section{Discussion and Conclusion}\label{SConclusion}

This paper provides a unified formalism describing the effects of SUSY breaking on the visible sector by extending the framework of GGM \cite{Meade:2008wd,Buican:2008ws,Komargodski:2008ax} to a large class of models where hidden-sector scalar chiral superfields couple to visible-sector scalar chiral superfields through the most general renormalizable superpotential.  Corrections to the visible-sector parameters are expressed in terms of one- and two-point correlation functions of hidden-sector fields.  Assuming a UV asymptotically-safe QFT, these hidden-sector correlation functions are approximated with the help of the OPE, which effectively disentangles the short-distance physics encoded in the OPE coefficients from the long-distance physics encoded in the vevs, leading to quantitative results which should be valid even in strongly-coupled hidden sectors.  Moreover, the perspective obtained from the OPE allows general comments to be made on the resulting phenomenology.

Indeed, the facts that the approximate results obtained from the OPE coefficients are order-one numbers times the exact results for a weakly-coupled example (see \cite{Fortin:2011ad}) and that non-perturbative physics are encoded in the vevs and thus completely disentangled from the OPE coefficients computed here suggest that the OPE techniques lead to reasonable approximations to visible-sector quantities even in strongly-coupled hidden theories.  Hence, barring unlikely cancellations between the different contributions to visible-sector quantities (generally fine-tuned unless some dynamical mechanism can explain them), generating natural visible-sector parameters with strongly-coupled theories necessitates to adequately control the relative sizes of the different vevs of the allowed operators appearing in the different OPEs.

The OPE techniques developed here are extensions of the works of \cite{Fortin:2011ad,Fortin:2012tp,Kumar:2014uxa}.  They are however more systematic since the computations of the OPE coefficients rely solely on the two- and three-point correlation functions.  As such, all conformal descendants are directly taken into account in the computations, instead of only a specific sample of conformal descendants needed to calculate OPE coefficients.

Finally, including to this work hidden-sector spinor superfields as well as Planck-suppressed non-renormalizable K\"ahler-potential and superpotential couplings between hidden-sector and visible-sector superfields would lead to the general framework of general gravity mediation.  The authors hope to return to such an idea in the future.


\ack{
The authors would like to thank Andreas Stergiou for enlightening discussions and comments on the manuscript.  This work is supported in parts by NSERC and FRQNT.
}


\bibliography{ScalarPortal}

\end{document}